\documentclass[12pt]{article}

\usepackage[left=1.0in,right=1.0in,top=1.0in,bottom=1.0in]{geometry}
\usepackage[version=3]{mhchem}
\usepackage[authoryear]{natbib}
\usepackage{fancyhdr}
\usepackage{xurl}
\usepackage{graphicx}
\usepackage{authblk}
\usepackage{gensymb}
\usepackage{array}

\title{\bf{A Bayesian hierarchical framework for fusion of remote sensing data: An example with solar-induced fluorescence}}

\author[1]{Manju Johny}
\author[1]{Jonathan Hobbs}
\author[1]{Vineet Yadav}
\author[1]{Margaret Johnson}
\author[1]{Nicholas Parazoo}
\author[1]{Hai Nguyen}
\author[1]{Amy Braverman}

\affil[1]{Jet Propulsion Laboratory, California Institute of Technology}

\date{February 15, 2025}

\makeatletter
\let\insertdate\@date
\makeatother

\begin{document}

\maketitle

\setlength{\headheight}{14pt}

\begin{abstract}
    Solar-induced chlorophyll fluorescence (SIF) has emerged as an effective indicator of vegetation productivity and plant health. The global quantification of SIF and its associated uncertainties yields many important capabilities, including improving carbon flux estimation, improving the identification of carbon sources and sinks, monitoring a variety of ecosystems, and evaluating carbon sequestration efforts. Long-term, regional-to-global scale monitoring is now feasible with the availability of SIF estimates from multiple Earth-observing satellites. These efforts can be aided by a rigorous accounting of the sources of uncertainty present in satellite SIF data products. In this paper, we introduce a Bayesian Hierarchical Model (BHM) for the estimation of SIF and associated uncertainties from Orbiting Carbon Observatory-2 (OCO-2) satellite observations at $1 \degree \times 1 \degree$ resolution with global coverage. The hierarchical structure of our modeling framework allows for convenient model specification, quantification of various sources of variation, and the incorporation of seasonal SIF information through Fourier terms in the regression model. The modeling framework leverages the predictable seasonality of SIF in most temperate land areas. The resulting data product complements existing atmospheric carbon dioxide estimates at the same spatio-temporal resolution.
\end{abstract}

\let\thefootnote\relax\footnotetext{\copyright 2025. All rights reserved.}

\clearpage

\section{Introduction}
 \label{sec:introduction} 

In recent years, solar-induced fluorescence (SIF) has emerged as an effective indicator of vegetation productivity and plant health. SIF, emitted by plants as red and near-infrared light in the 650-850 nm spectral range during photosynthesis, serves as a proxy for vegetation productivity \citep{kohler_tropomi_2020}. While traditional greenness-based vegetation indices, such as the normalized difference vegetation index (NDVI), have a slower response to plant health, SIF serves as a near immediate indicator of plant health and productivity due to its close tie to the photosynthetic process \citep{frankenberg_grl_2011, xu_2021}. Furthermore, recent advances in global SIF retrievals through satellite instruments have led to the increased availability of broad spatio-temporal information related to plant photosynthesis, previously infeasible through in situ and airborne measurements alone \citep{sif_retrieval_yao}.

Global observations of SIF have been facilitated by several satellites, including the Greenhouse Gases Observing Satellite \citep[GOSAT;][]{frankenberg_grl_2011}, Global Ozone Monitoring Experiment-2 \citep[GOME-2;][]{joiner_global_2013}, SCanning Imaging Absorption spectroMeter for Atmospheric CartograpHY \citep[SCIAMACHY;][]{joiner_sciamachy_2021}, Orbiting Carbon Observatory-2 \citep[OCO-2;][]{sun_oco2_2018}, Orbiting Carbon Observatory-3 \citep[OCO-3;][]{doughty_ocosif_2022}, and TROPOspheric Monitoring Instrument \citep[TROPOMI;][]{kohler_tropomi_2018}. \citet{parazoo_jgr_2019} provide an overview and comparisons among GOME-2, OCO-2, and TROPOMI SIF. A common retrieval framework for SIF using microwindows along the edges of the oxygen A-band spectra has been implemented for GOSAT, OCO-2, and OCO-3; an outline of the SIF products produced by these platforms is provided by \citet{doughty_ocosif_2022}. Among these, OCO-2 provides the finest spatial resolution (1.3 km $\times$ 2.25 km), with a relatively long time span of measurements collected from 2014 to present. In addition, greenhouse gas monitoring missions that provide coincident SIF estimates offer a unique opportunity for combined carbon cycle inference \citep{zhang_sifinv_2023}. \citet{jacobson_spatial_2023} use OCO-2 data to develop coSIF, which uses cokriging to predict SIF by using the lagged spatiotemporal dependence between the mole fraction of carbon dioxide (XCO$_2$) and SIF at a $0.05 \degree$  target resolution. 

There have also been a number of aggregated SIF products based on satellite measurements of SIF. Among these are the global OCO-2 SIF dataset \citep[GOSIF;][]{li_gosif_2019} and contiguous solar-induced fluorescence \citep[CSIF;][]{zhang_csif_2018}, which integrate measurements of OCO-2 SIF with Moderate Resolution Imaging Spectroradiometer (MODIS), to provide fine resolution global SIF products at $0.05 \degree$ degree resolution. Due to the relatively small target resolutions achieved by these products, they have been useful for ecosystem to regional-scale studies. Gross Primary Productivity (GPP) represents the total amount of carbon dioxide utilized by plants through photosynthesis. Reliable estimates of GPP are valuable for a variety of applications, including studies investigating carbon cycle-climate interactions and feedbacks \citep{beer_terrestrial_2010}. Satellite measurements of SIF have emerged as a reliable indicator of GPP, enabling regional and global quantification of primary productivity \citep{mohammed_remote_2019,parazoo_jgr_2019}. Due 
to the crucial role of photosynthesis within the carbon cycle and its close correlation with GPP, SIF has been a useful constraint in carbon cycle data assimilation systems (CCDAS), such as BETHY-SCOPE \citep{bethy-scope}, CARDAMOM \citep{cardamom}, and ORCHIDAS \citep{orchidas}. Additionally, SIF has been useful for informing carbon dioxide fluxes and net ecosystem exchange (NEE) at regional scales \citep{shiga_2017,zhang_sifinv_2023}.

In this paper, we introduce a gridded SIF product based on Bayesian Hierarchical Modeling (BHM), which provides SIF estimates with quantified uncertainties at $1 \degree \times 1 \degree$ gridded resolution for global, terrestrial regions based on retrievals of SIF from the OCO-2 satellite. The model utilizes OCO-2 SIF retrievals along with seasonal information derived from TROPOMI SIF to obtain daily estimates of SIF for days with available OCO-2 SIF data. By relying only on SIF measurements, we hope to reduce biases caused by the incorporation of other vegetation measures, such as NDVI. Satellite retrievals of SIF are subject to multiple sources of uncertainty \citep{frankenberg_grl_2011,kohler_tropomi_2018}. Our methodology employs a hierarchical statistical model to provide a comprehensive probabilistic framework to account for these sources and the relationships among them. Combined with a Bayesian formalism, the hierarchical model provides a convenient framework for inference, including estimation and uncertainty quantification (UQ), for an unobserved target process at a desired spatio-temporal resolution, given irregular and diverse observations \citep{berliner_hierarchical_1996}. In addition to accounting for satellite retrieval uncertainty, the BHM serves as a convenient framework for partitioning spatio-temporal variability in SIF. In particular, the BHM in this work separates local spatial variability from seasonal evolution of the mean SIF into different levels of the model, while the Bayesian formulation of the model allows scientific expertise to inform the seasonal cycle. This mathematical framework for spatially-varying seasonality in the carbon cycle has been adopted previously for atmospheric CO$_2$ concentrations \citep{lindqvist_2015} and for CO$_2$ fluxes \citep{bertolacci_wombatv2_2024}.

The BHM gridded SIF are co-located with the OCO-2-based XCO$_2$ estimates and their associated uncertainties obtained by local kriging, as described in \citet{nguyen_meas_atbd_v4}, and provided in a combined SIF and XCO$_2$ product \citep{measures_v4_sif}. The combined product allows joint inference of SIF and XCO$_2$, as well as the ease of integration of these variables into carbon flux and NEE inversion algorithms, as well as CCDAS. This facilitates many important capabilities including robust quantification of carbon budgets, identification of carbon sources and sinks, assessment of carbon sequestration efforts, and more robust understanding of the biosphere. While our modeling framework is flexible with regard to the target spatial resolution, we set our target resolution to $1 \degree$ gridded regions to be readily usable with the co-located XCO$_2$ in global flux inversion algorithms, which are limited to resolutions between $1 \degree$ and $7 \degree$ \citep{crowell_2019}. 

The remainder of this paper is organized as follows. The data sources are discussed in Section \ref{sec:data}. The hierarchical statistical model structure, implementation, and outputs are outlined in Section 3. Example results and comparisons to other products are summarized in Section 4. Further discussion is included in Section 5, and concluding remarks are provided in Section 6.

\section{Data} \label{sec:data}

This section describes multiple input data sources that are used in the development and illustration of the BHM gridded SIF product. We first describe two static global ecological data products that provide contextual information for the estimation and analysis. This is followed by details of the OCO-2 and TROPOMI SIF products used as inputs to the BHM.

\subsection{MODIS Land Cover (MCD12C1) Version 6} \label{sec:data-modis}
The Terra and Aqua combined MODIS Land Cover Climate Modeling Grid (CMG) (MCD12C1) Version 6 data product \citep{MCD12C1} provides the majority land cover classifications at 0.05$\degree$ $\times$ 0.05$\degree$ resolution, with global coverage annually from 2001 to 2020. The classification types are provided in Table~\ref{land-cover-table}. The majority land cover map for 2019 is up-scaled to 1$\degree$ $\times$ 1$\degree$ resolution by obtaining the mode of the majority land cover classifications within each grid cell, excluding water bodies. The majority land cover classification maps are used during pre-processing of the OCO-2 SIF data, as described in Section \ref{sec:data-oco2}. 

\begin{table}[h]
\begin{center}
\caption{MODIS MCD12C1 Version 6 majority land cover type with corresponding values.}
\label{land-cover-table}
\vspace*{0.15in}
\begin{tabular}{ | l l |} 
  \hline
  \textbf{Value} & \textbf{Name} \\
  \hline
    1 &  Evergreen Needleleaf Forests  \\ \hline
    2 &  Evergreen Broadleaf Forests \\ \hline
    3 &  Deciduous Needleleaf Forests \\ \hline
    4 &  Deciduous Broadleaf Forests \\ \hline
    5 &  Mixed Forests \\ \hline
    6 &  Closed Shrublands \\ \hline
    7 &  Open Shrublands \\ \hline
    8 &  Woody Savannas  \\ \hline
    9 &  Savannas \\ \hline
    10 & Grasslands  \\ \hline
    11 & Permanent Wetlands  \\ \hline
    12 & Croplands  \\ \hline
    13 & Urban and Built-up Lands   \\ \hline
    14 & Cropland/Natural Vegetation Mosaics  \\ \hline
    15 & Permanent Snow and Ice   \\ \hline
    16 & Barren   \\ \hline
    17 & Water Bodies   \\ \hline
    255 & Unclassified   \\ \hline
\end{tabular}
\end{center}
\end{table} 

\subsection{RESOLVE Ecoregions 2017} \label{sec:data-biome}
The Ecoregions 2017 dataset \citep{biome} gives a map of 846 terrestrial ecoregions, grouped into biomes. The BHM gridded SIF are grouped into biome types, given in Table~\ref{biome-table}, and subsequently used for descriptive analysis as described in Section \ref{sec:results-results}, and to establish comparisons with TROPOMI SIF as described in Section \ref{sec:results-comparisons}.

\begin{table}[h]
\begin{center}
\caption{RESOLVE Ecoregions 2017 biome type with corresponding values.}
\label{biome-table}
\vspace*{0.15in}
\begin{tabular}{ | l l |} 
  \hline
  \textbf{Value} & \textbf{Name} \\
  \hline
    1 &  Tropical \& Subtropical Moist Broadleaf Forests  \\ \hline
    2 &  Tropical \& Subtropical Dry Broadleaf Forests \\ \hline
    3 &  Tropical \& Subtropical Coniferous Forests \\ \hline
    4 &  Temperate Broadleaf \& Mixed Forests \\ \hline
    5 &  Temperate Conifer Forests  \\ \hline
    6 &  Boreal Forests/Taiga  \\ \hline
    7 &  Tropical \& Subtropical Grasslands, Savannas \& Shrublands \\ \hline
    8 &  Temperate Grasslands, Savannas \& Shrublands  \\ \hline
    9 &  Flooded Grasslands \& Savannas \\ \hline
    10 & Montane Grasslands \& Shrublands   \\ \hline
    11 & Tundra \\ \hline
    12 & Mediterranean Forests, Woodlands \& Scrub  \\ \hline
    13 & Deserts \& Xeric Shrublands   \\ \hline
    14 & Cropland/Natural Vegetation Mosaics  \\ \hline
    15 & Mangroves    \\ \hline
\end{tabular}
\end{center}
\end{table}

\subsection{OCO-2 Level 2 SIF, version 11} \label{sec:data-oco2}
 OCO-2 SIF data from the version 11 ``Lite'' SIF product \citep{ltsif_v11} was used as the data source for the BHM SIF estimation model described in Section \ref{sec:methods}. The data are provided for native OCO-2 footprints, which have a spatial resolution of approximately 2.25 km $\times$ 1.3 km, and temporal coverage from 2014-09-06 to present. The OCO-2 polar orbit has a revisit frequency of 16 days. As discussed in \citet{doughty_ocosif_2022}, the OCO-2 SIF products contain quality flags for each observation with 3 levels: Best (0), Good (1), and Failed (2). We use only the Best (0) and Good (1) quality flagged data. Ocean observations are removed using the MCD12C1 (0.05 Deg) majority land cover classifications. Additionally, the up-scaled ($1\degree$) majority land cover classification, described in Section \ref{sec:data-modis}, are used to remove $1 \degree$ grid cells classified as (15) Permanent Snow and Ice, (16) Barren, and (17) Water Bodies from the BHM estimation pipeline due to negligible expected photosynthetic activity in these regions. After the above pre-processing, OCO-2 SIF from the years 2014 to 2023 are used to obtain BHM gridded SIF estimates.
 
\subsection{TROPOspheric Monitoring Instrument (TROPOMI) SIF} \label{sec:data-tropomi}
TROPOMI, on board the Copernicus Sentinel-5P mission, provides SIF data at a spatial resolution of 3.5 km $\times$ 5.5 km, daily revist, and temporal coverage from March 2018 to present \citep{tropomi_dataset_caltech, kohler_tropomi_2018}. The daily revisit frequency of TROPOMI provides a more temporally dense SIF dataset than OCO-2 within a year, allowing it to capture finer seasonal dynamics of the SIF. We utilize TROPOMI SIF from the years 2020 and 2021 to inform a prior distribution for the SIF seasonal cycle in the BHM, as detailed in Section \ref{sec:methods}. The TROPOMI SIF from 2019  was left out of the estimation pipeline, and is used instead to establish comparisons with the BHM gridded SIF in Section \ref{sec:results-comparisons}. 

\section{Methods}
\label{sec:methods} 

\subsection{Model Rationale} \label{sec:methods-model-rationale} 
We obtain gridded SIF estimates for $1 \degree \times 1 \degree$ regions globally using OCO-2 SIF data. The OCO-2 SIF data are first grouped by grid cell, then grouped by day, to obtain a collection of SIF observations that can be used to obtain daily, gridded SIF estimates. This collection of SIF observations can be considered as realizations from a latent SIF process, observed at sounding locations, and measured with error. Our primary objective is to model this latent SIF process for each grid cell and day, to obtain its estimates as the daily, gridded SIF product. A BHM allows for this kind of grouping structure and offers several additional advantages: (1) The ability to incorporate measurement error, (2) The ability to separate and quantify various sources of variation, and (3) The ability to obtain estimates of the process or parameters of interest with quantified uncertainty. 

\subsection{Model Specification} \label{sec:methods-model-specification} 

We specify a Bayesian hierarchical regression model to obtain the BHM gridded SIF estimates. Modeling for each year and grid cell is done independently for the years 2014 through 2023.

A BHM breaks a potentially complex joint probability distribution into three stages of conditional distributions \citep{berliner_hierarchical_1996,cressie_wikle}. The stages can be summarized as follows, where $[A]$ represents the distribution of $A$ and $[A|B]$ represents the conditional distribution of $A$ given $B$. 
\begin{enumerate}
    \item Data Model: $[data|process, parameters]$
    \item Process Model: $[process|parameters]$
    \item Parameter Model: $[parameters]$
\end{enumerate}
We infuse prior knowledge about the parameters in the parameter model stage, and use Bayes' theorem to obtain posterior estimates of the process and parameters of interest, conditioned on the observed data. The posterior distribution can be obtained from the three stages by the following relationship:
\vspace{-0.1in}
\begin{equation}
\begin{split}
     [process, parameters | data] \propto [data|process, parameters][process|parameters][parameters]
\end{split}
\end{equation}
The general BHM framework facilitates the following specification for the SIF remote sensing products.
Within a grid cell, for an OCO-2 overpass on day of year (Julian day) $t$, let $Z_{it}$ represent the $i^{th}$ sounding's retrieval of SIF from OCO-2. The data model is written as
\begin{eqnarray}
Z_{it} &=& Y_{it} + m_{it}, \hspace{0.5in} i=1, \ldots n_t,
\label{eq:data_mod_z}
\end{eqnarray}
where $n_t$ is the total number of retrievals in the grid cell on day $t$.
In the formulation (\ref{eq:data_mod_z}), $Y_{it}$ represents the noise-free, latent SIF at each $i^{th}$ sounding location, which can only be observed with measurement error, represented by $m_{it}$. We specify a distribution for these errors, $m_{it} \overset{iid}{\sim} N(0, \tau_{it})$ where $\tau_{it}$ represents the OCO-2 SIF data product retrieval error variance obtained from the OCO-2 Lite files, as estimated using the procedure described in \citet{sun_oco2_2018}.

In the process model stage, we represent complex spatio-temporal structures of the latent SIF process, which can be broken into two submodels. In the first process submodel, for each day, we represent deviations of the latent, sounding-level SIF process ($Y_{it}$) from the latent, mean SIF process ($X_t$) at $1 \degree \times 1 \degree$ resolution.
\begin{eqnarray}
Y_{it} &=& X_t + r_{it} 
\label{eq:proc_mod_y}
\end{eqnarray}
In the above formulation, $r_{it}$ represents deviations of the sounding-level SIF processes from the $1 \degree$ mean SIF process, which can be thought of as the small-scale spatial variability of the SIF process within the grid cell. Estimates of $X_t$ for each grid cell on day $t$, along with their standard errors, comprise the BHM gridded SIF product. 

In the second process submodel, we model the latent, $1 \degree$ mean SIF in terms of its underlying seasonal behavior. The photosynthetic activity of a region rises and falls according to vegetation seasonal cycles. We utilize Fourier terms to represent this cyclical behavior of the SIF process over time. This model allows for a concise and interpretable representation of vegetation seasonality, which can be utilized to inform the $1 \degree$ mean SIF, $X_t$, and decrease its uncertainty. 
\begin{eqnarray}
X_{t} &=& \mu_t + d_t \label{eq:proc_mod_x}
 \\
\mu_t &=& a+ \beta_0 + \beta_1 t + \sum_{k=1}^{K} \Bigg [ \beta_{2,k} sin \bigg(\frac{2k\pi t}{365.25}\bigg) + \beta_{3,k} cos \bigg(\frac{2k\pi t}{365.25}\bigg) \Bigg ] \label{eq:proc_mod_mu}
\end{eqnarray}
In (\ref{eq:proc_mod_mu}), $\mu_t$ represents the underlying seasonality of SIF, specified by an intercept term, trend term, and Fourier terms to capture cyclical patterns over time. We truncate the number of Fourier terms, $K=2$, due to temporal sparsity in the OCO-2 data. For individual overpasses, $d_t$ represents deviations of the $1 \degree$ mean SIF process, $X_t$, from its underlying seasonal structure. An additional vertical shift term, $a$, gives added flexibility to the seasonal cycle to adjust according to the magnitude of SIF. Incorporating a smoothly varying seasonal structure in the process model stage is a common practice for hierarchical statistical models for environmental time series \citep{kleiber_daily_2013,poppick_quantreg_2020}.

Lastly, we specify distributions for the error terms in our process model. For the $i^{th}$ sounding on day $t$,
\begin{itemize}
    \item[] $r_{it}\overset{iid}{\sim} N (0, \nu_t) $ where $\nu_t$ is to be estimated. 
    \item[] $d_t\overset{iid}{\sim} N (0, \delta) $ where $\delta$ is to be estimated. 
\end{itemize}

The variance parameters, $\nu_t$, describe the expected spatial variability of sounding-level SIF around the $1\degree$ mean SIF, $X_t$, while $\delta$ describes the expected intraseasonal temporal variability of the $1 \degree$ mean SIF from the smooth seasonality term, $\mu_t$.

In the parameter stage, we specify prior distributions for the unknown parameters. We utilize `vague' priors in the cases where enough information is not known about the distributions a priori \citep{cressie_wikle}. We specify Exp(1) distributions for $\frac{1}{\nu_t}$ and $\frac{1}{\delta}$, which gives positive and disperse support over the parameter space. We specify a Unif(-1,1) prior distribution for the vertical shift term, $a$, in the seasonal cycle to give flexibility for the seasonal cycle to adjust according to the magnitude of SIF for a particular year. For the $\beta_0, \beta_1, \beta_{2,k}, \beta_{3,k}$ coefficients of the seasonal cycle, we specify informative priors obtained from TROPOMI SIF. We specify the prior distributions of the beta coefficients as 
\begin{itemize}
    \item[] $\beta_0 \sim N(b_0, s_0)$
    \item[] $\beta_1 \sim N(b_1, s_1)$
    \item[] $\beta_{2,k} \sim N(b_{2,k}, s_{2,k})$
    \item[] $\beta_{3,k} \sim N(b_{3,k}, s_{3,k})$
\end{itemize}
where the $b,s$ prior parameters are obtained using TROPOMI SIF. A Bayesian hierarchical regression model of the same formulation, with Unif(-1,1) priors on the $\beta$ coefficients, are modeled on combined TROPOMI SIF data from 2020 and 2021 to obtain hyperparameters, $b$ representing the posterior mean of the coefficients and $s$ representing the posterior variance of the coefficients. These estimates are subsequently fed into our model as prior hyperparameters for the seasonal cycle. Utilizing TROPOMI-informed hyperparameters allows us to infuse seasonal information obtained from the temporally dense TROPOMI data, while allowing sufficient flexibility for the seasonal cycle to update according to the OCO-2 SIF for a particular year.

\subsection{Implementation} \label{sec:methods-implementation} 
A Markov Chain Monte Carlo (MCMC) algorithm was used to obtain posterior distributions for the model parameters and derive quantities of interest. The MCMC was run using a Gibbs sampling algorithm, implemented through the \texttt{JAGS} software \citep{jags_manual} and the \texttt{R2jags} package \citep{r2jags} in the \texttt{R} programming language. The SIF estimation for each grid cell is performed for a full year, for the years 2014 through 2023 independently. Further, the model specification allows each grid cell to be modeled independently, allowing the MCMC for the grid cells to be run embarrassingly parallel. 

\subsection{Output Variables} \label{sec:methods-output-variables} 
The output variables for the BHM gridded SIF product are summarized in Table~\ref{BHM-variable-table}. We give further details about the calculation of these variables below. The SIF estimates in the BHM gridded SIF product are given by the variable, sif\textunderscore740nm. It is obtained as the posterior mean of the $1 \degree$ SIF process, $X_{t}$, for each grid cell on day $t$. The uncertainty of the SIF estimate, sif\textunderscore uncertainty, is calculated as the standard deviation of the posterior distribution of the $1 \degree$ SIF process. The posterior $2.5^{th}$ and $97.5^{th}$ quantiles of the gridded SIF are given by the variables sif\textunderscore quantile\textunderscore2.5 and sif\textunderscore quantile\textunderscore97.5. They are obtained as quantiles of the posterior distribution of the $1 \degree$ SIF process, and together give the 95$\%$ credible interval for the BHM gridded SIF. 

The summaries of the posterior distribution of the BHM gridded SIF are supplemented by other contextual information to aid in the interpretation of the SIF estimates. For each grid cell, The majority land cover, sif\textunderscore land\textunderscore cover, was obtained using the MODIS Land Cover (MCD12C1) data as described in Section \ref{sec:data-modis}. Since we do not expect rapid land cover changes at the $1 \degree$ product resolution, the 2019 MCD12C1 data was used to obtain the majority land cover variable for all years. The levels of the sif\textunderscore land\textunderscore cover variable are consistent with the MCD12C1 levels, and are summarized in Table~\ref{land-cover-table}. We further note that grid cells classified as (15) `Permanent Snow and Ice', (16) `Barren', and (0) `Water Bodies' were removed from the processing pipeline, and not included in the BHM gridded SIF product.

The time variable, sif\textunderscore time, is given in seconds since 1970-01-01 00:00:00. When corresponding daily estimates of XCO$_2$ are available for a grid cell, we use the same time indices as the XCO$_2$ counterpart, obtained as the kriging-weighted average of the sounding times obtained from XCO$_2$ sounding times \citep{measures_v4_sif}. This is to allow for as many coincident measurements between the XCO$_2$ and SIF as possible within the combined product. When corresponding daily estimates of XCO$_2$ are unavailable for the grid cell, the time variable is calculated as a simple average of the OCO-2 SIF sounding times.

\begin{table}[h]
\begin{center}
\caption{Names and descriptions of variables included in the Bayesian Hierarchical Model (BHM) gridded SIF product.}
\label{BHM-variable-table}
\vspace*{0.15in}
\begin{tabular}{ | m{0.18\linewidth}  m{0.24\linewidth}  m{0.54\linewidth}| } 
  \hline
  \textbf{Variable Name} & \textbf{Units} & \textbf{Description} \\ 
  \hline
  sif\textunderscore 740nm & W m$^{-2}$ sr$^{-1}$ $\mu$m$^{-1}$ & Daily, gridded estimates of solar-induced chlorophyll fluorescence (SIF) at 740 nm \\ 
  \hline
  sif\textunderscore uncertainty & W m$^{-2}$ sr$^{-1}$ $\mu$m$^{-1}$ & Uncertainty (standard error) of the daily, gridded estimates of solar-induced chlorophyll fluorescence (SIF) at 740 nm   \\ 
  \hline
  sif\textunderscore quantile\textunderscore2.5 & W m$^{-2}$ sr$^{-1}$ $\mu$m$^{-1}$ & $2.5^{th}$ quantile of the daily, gridded estimates of solar-induced chlorophyll fluorescence (SIF) at 740 nm  \\ 
  \hline
  sif\textunderscore quantile\textunderscore97.5 & W m$^{-2}$ sr$^{-1}$ $\mu$m$^{-1}$ & $97.5^{th}$ quantile of the daily, gridded estimates of solar-induced chlorophyll fluorescence (SIF) at 740 nm  \\ 
  \hline
  sif\textunderscore land\textunderscore cover & N/A &  Majority land cover classifications ($1\degree$) with levels as described in Table \ref{land-cover-table}. \\ 
  \hline
  sif\textunderscore latitude & Degrees North & Center latitude corresponding to sif\textunderscore nm  \\ 
  \hline
  sif\textunderscore longitude &  Degrees East & Center longitude corresponding to sif\textunderscore nm   \\ 
  \hline
  sif\textunderscore time & Seconds & Time corresponding to sif\textunderscore 740nm, reported as seconds since 1970-01-01 00:00:00  \\ 
  \hline
  sif\textunderscore date & N/A & Two-dimensional array storing the full date and time corresponding to each observation of sif\textunderscore 740nm; organized as year, month, day, hour, minute, second, milliseconds.  \\ 
  \hline
\end{tabular}
\end{center}
\end{table}

\section{Results and Comparisons} \label{sec:results} 
\subsection{Results} \label{sec:results-results} 
\begin{figure}[h]
\centering
\rotatebox[origin=c]{90}{\includegraphics[width=1.1\linewidth]{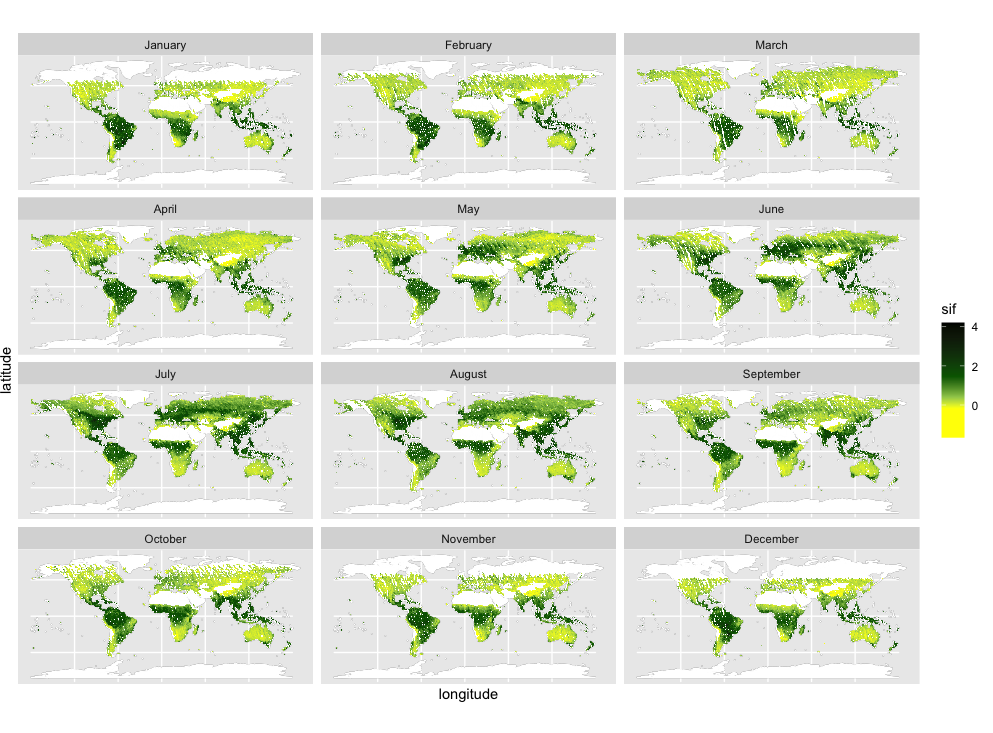}}
\caption{Global map of the monthly average SIF obtained from the BHM gridded SIF for the year 2019. SIF units are W m$^{-2}$ sr$^{-1}$ $\mu$m$^{-1}$.}
\label{fig:global_sif_map_app}
\end{figure}

\begin{figure}[h]
\centering
\includegraphics[width=1\linewidth]{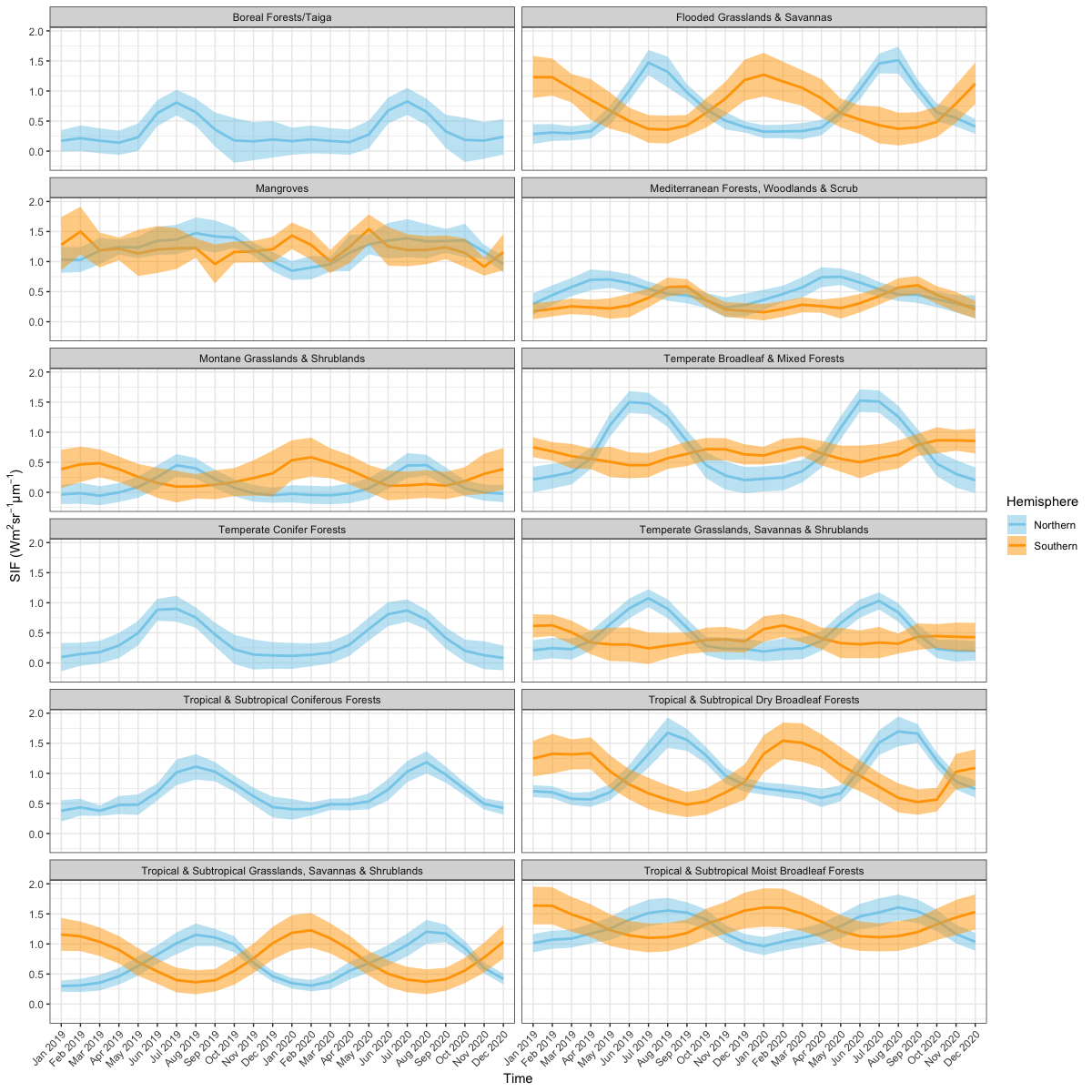}
\caption{The mean SIF (solid line) with 95$\%$ credible intervals (ribbon) are shown for the BHM gridded SIF, aggregated by month for each biome, for 2019 through 2020. Blue and orange indicate SIF for the northern and southern hemispheres respectively.}
\label{fig:bhm_mean_sif_95p}
\end{figure}

The BHM gridded SIF product contains global, gridded SIF estimates along with associated measures of uncertainty, the standard error of the estimated SIF and the $95\%$ credible interval of the estimated SIF, for the years 2014 through 2023. The gridded SIF estimates are obtained daily for days with available OCO-2 data. We illustrate the product using the example years 2019 and 2020. Figure \ref{fig:global_sif_map_app} shows global maps of the monthly averaged SIF calculated from the BHM gridded SIF for the year 2019. In the northern hemisphere, the spring months (March, April, May) have relatively low values of SIF that gradually increase towards the summer months (June, July, August). The summer months have high values of SIF, particularly in the US corn belt, central Europe, and northern China. From autumn (September, October, November) through winter (December, January, February), we observe SIF values declining as the growing cycle winds down. Similar seasonal patterns in the mean SIF and uncertainties is observed across biomes in the northern hemisphere (Figure \ref{fig:bhm_mean_sif_95p}). We notice a pronounced seasonal cycle in the northern hemisphere across most biomes, with strong peaks in the summer or early autumn. In the southern hemisphere, particularly in southern portions of South America and Africa, we observe relatively low SIF during the austral winter months (June, July, August), and high SIF during the austral summer months (December, January, February). Similar seasonal patterns in the mean SIF and uncertainties can be observed across biomes in the southern hemisphere, with high SIF values observed early and late in the year, and a dip in SIF around July and August (Figure \ref{fig:bhm_mean_sif_95p}). 

\subsection{Comparison of BHM Gridded SIF with TROPOMI SIF} \label{sec:results-comparisons} 

TROPOMI SIF provides the most temporally dense measurements of SIF available at a global scale since 2018. The daily revisit frequency of TROPOMI allows for a more detailed understanding of the seasonal trends of SIF globally compared to OCO-2, which has a revisit frequency of 16 days. In this section, we establish comparisons of the seasonal trends observed in the BHM gridded SIF with those observed in the TROPOMI SIF. Comparisons are established in the biome regions given in Table \ref{biome-table} for the year 2019. We note that the prior distributions for the seasonal coefficients in the BHM were derived only from 2020 and 2021 TROPOMI SIF, and 2019 TROPOMI SIF was left out of the BHM estimation pipeline. The comparison methodology is described below. 
\begin{enumerate}
    \item We compare the distribution of SIF between the BHM gridded SIF and the TROPOMI SIF for the year 2019 for each biome. The BHM gridded SIF is aggregated to monthly temporal scales by obtaining the monthly average SIF for each grid cell within a biome. The TROPOMI SIF is first aggregated to $1 \degree \times 1\degree$ resolution, and subsequently aggregated monthly within a biome and grid cell. We obtain box plots of SIF by month of the two products to produce biome-level graphical comparisons of the $1 \degree$ SIF distributions. Separate figures are obtained for the northern and southern hemispheres to account for the different seasonal cycles between the hemispheres. 
    \item We compare the mean SIF and the associated uncertainties between the BHM gridded SIF and the TROPOMI SIF for the year 2019 for each biome. Within a biome, the monthly average SIF and corresponding uncertainty (standard deviation scale) are obtained for the BHM gridded SIF and the TROPOMI SIF. Line and ribbon plots provide biome-level comparisons of the mean SIF and corresponding uncertainties between each product. Separate figures are obtained for the northern and southern hemispheres to account for the different seasonal cycles between the hemispheres.  
\end{enumerate}

\begin{figure}[h]
\centering
\includegraphics[width=1\linewidth]{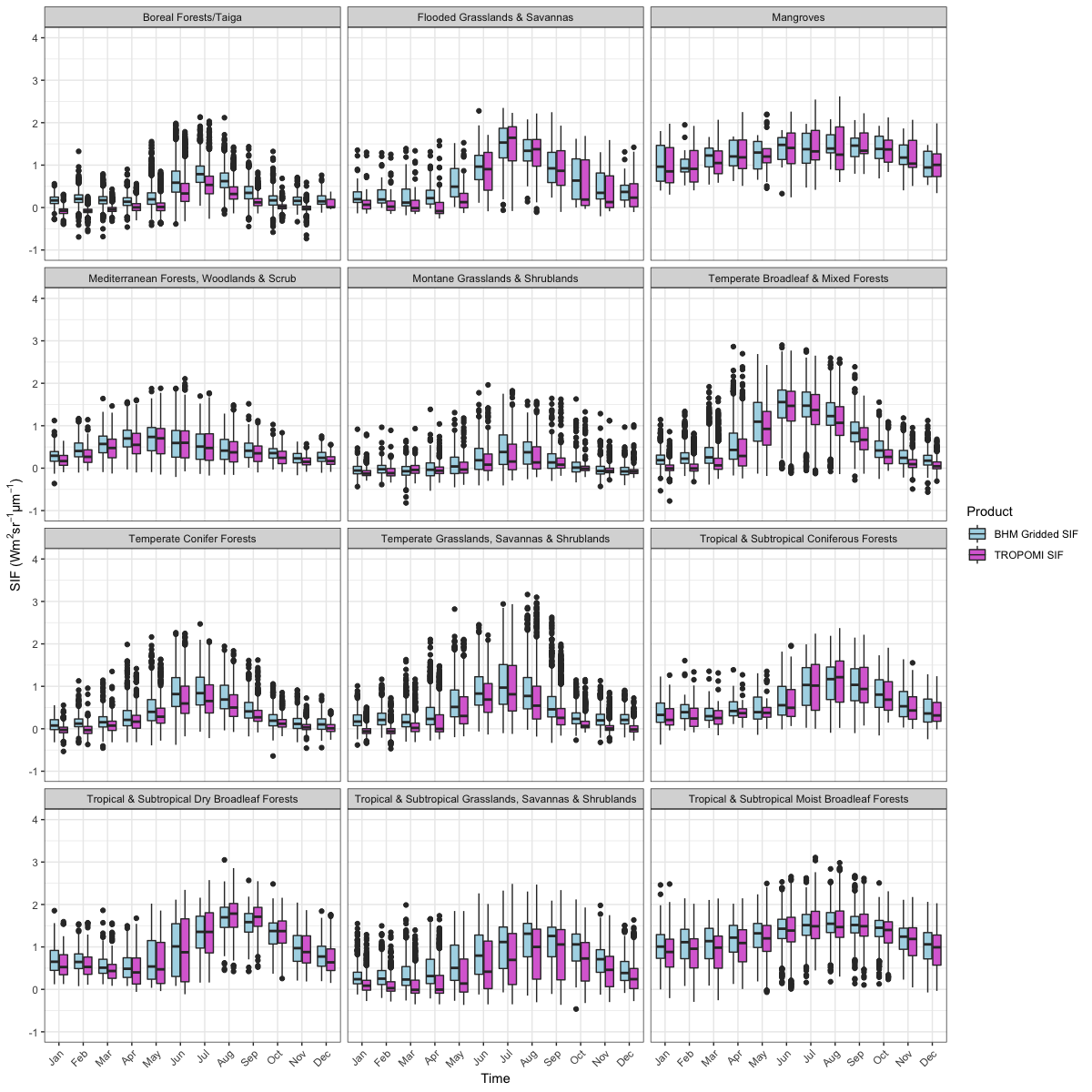}
\caption{Comparison of the distribution of SIF between aggregated TROPOMI SIF (pink) and the BHM gridded SIF product (blue) for each biome in the northern hemisphere.}
\label{fig:boxplots_north}
\end{figure}

\begin{figure}[h]
\centering
\includegraphics[width=1\linewidth]{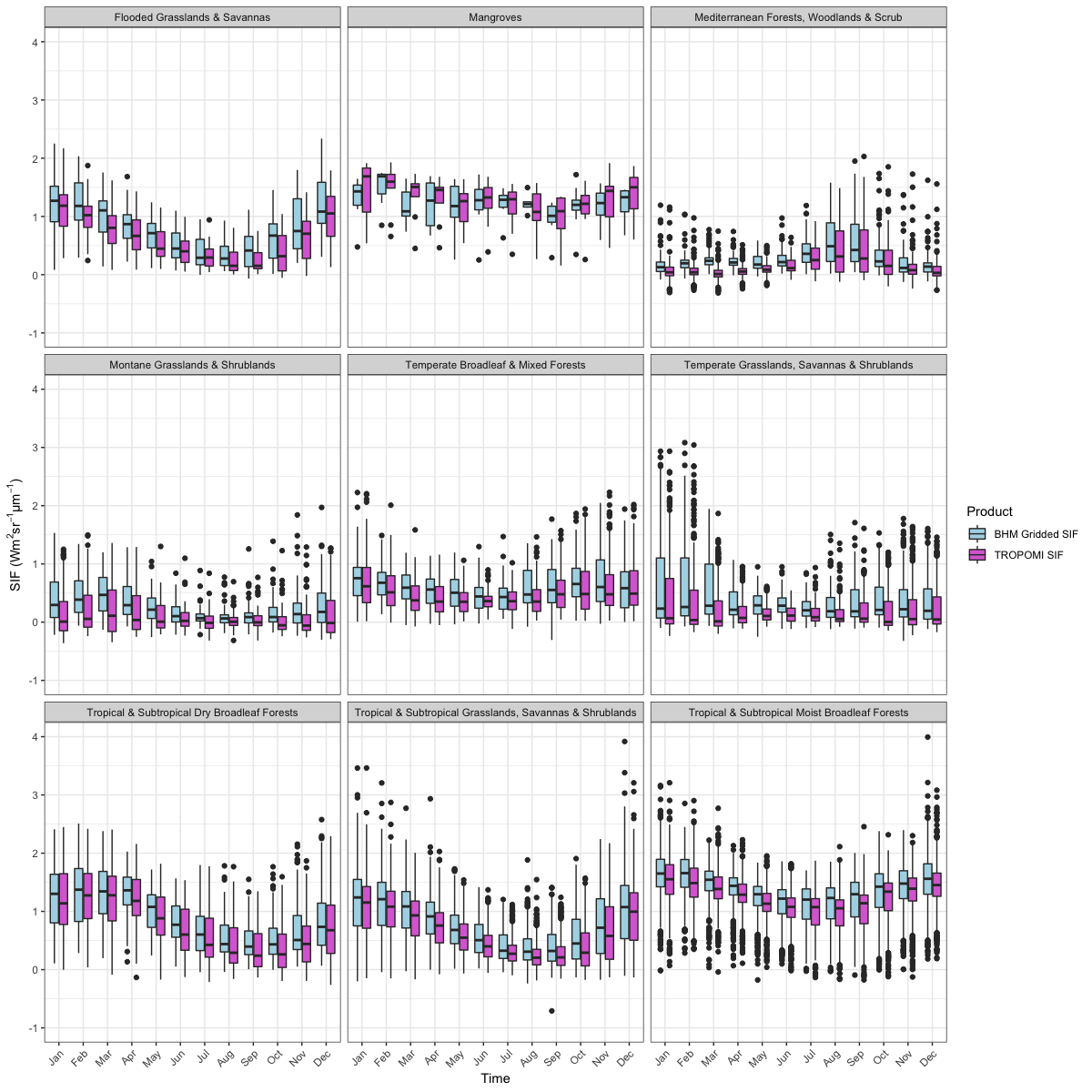}
\caption{Comparison of the distribution of SIF between aggregated TROPOMI SIF (pink) and the BHM gridded SIF product (blue) for each biome in the southern hemisphere is shown.}
\label{fig:boxplots_south}
\end{figure}

\begin{figure}[h]
\centering
\includegraphics[width=1\linewidth]{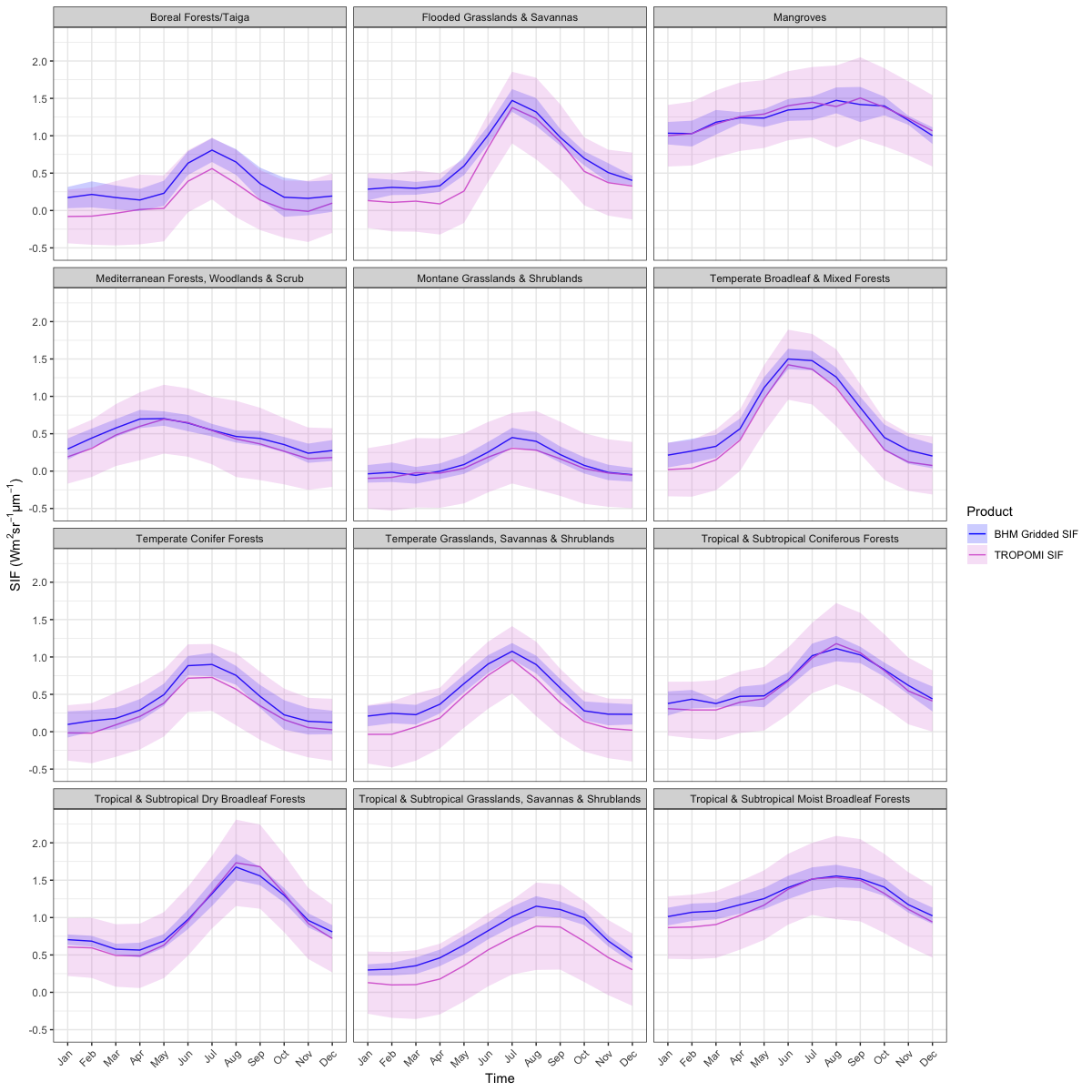}
\caption{Comparison of average monthly SIF along with their uncertainties bounds ($\pm$1 standard deviation) between TROPOMI SIF (pink) and the BHM gridded SIF product (blue) for each biome in the northern hemisphere.}
\label{fig:sif_uncert_north}
\end{figure}

\begin{figure}[h]
\centering
\includegraphics[width=1\linewidth]{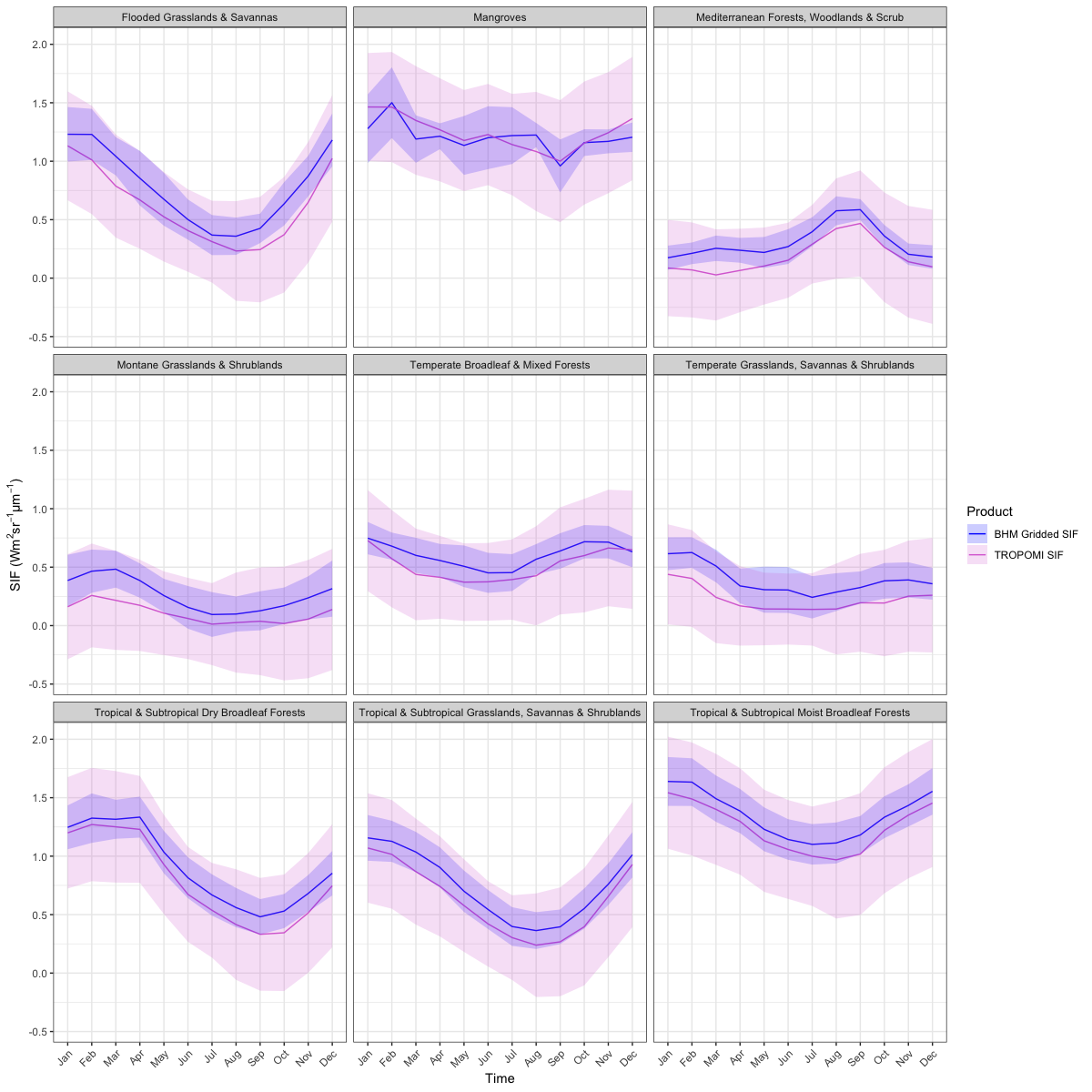}
\caption{Comparison of average monthly SIF along with their uncertainties bounds ($\pm$ 1 standard deviation) between TROPOMI SIF (pink) and the BHM gridded SIF product (blue) for each biome in the northern hemisphere.}
\label{fig:sif_uncert_south}
\end{figure}

 In Figures \ref{fig:boxplots_north} and \ref{fig:boxplots_south}, we compare the distribution of monthly averaged SIF between the BHM gridded SIF and TROPOMI SIF for the northern and southern hemispheres respectively. Similar seasonal patterns across biomes are observed between the two products in the northern hemisphere (Figure \ref{fig:boxplots_north}) and southern hemisphere (Figure \ref{fig:boxplots_south}). In Figures \ref{fig:sif_uncert_north} and \ref{fig:sif_uncert_south}, we compare the mean monthly SIF values along with their uncertainties ($\pm 1$ standard deviation) between the BHM gridded SIF and TROPOMI SIF. The TROPOMI uncertainties used for the figures were calculated by first aggregating to $1\degree$, then aggregating by month and biome. The BHM gridded SIF uncertainties are obtained as described in Section \ref{sec:methods}, and further aggregated by biome and month. We observe overlap between the mean SIF and uncertainty bounds for TROPOMI and the BHM gridded SIF products, with BHM gridded SIF exhibiting narrower uncertainty bounds across all biomes.

\clearpage
\section{Discussion} \label{sec:discussion} 
The global maps and boxplots, illustrating the distribution of the BHM gridded SIF, show agreement of the SIF estimates with expected spatial patterns of vegetation productivity globally. In particular, we note that the BHM SIF are able to capture seasonal trends of the SIF, and show distinct timing of seasonal peaks in the northern and southern hemispheres in all biomes. Further, the distribution of SIF, plotted as a time series in Figure 2, illustrates the continuity of the SIF product year to year, with consistent seasonal patterns observed in 2019 and 2020. 

Furthermore, comparisons of the BHM gridded SIF with TROPOMI SIF yield promising results, indicating good agreement between the products in terms of both the seasonal and spatial patterns of SIF. The BHM SIF exhibits narrower uncertainty intervals around the mean SIF (Figure \ref{fig:sif_uncert_north} and \ref{fig:sif_uncert_south}) for each biome compared to TROPOMI SIF. There are several possible reasons for this. First, we note that the uncertainty intervals obtained from the BHM product are calculated from the standard errors of the gridded SIF estimates, while the uncertainty intervals for TROPOMI are obtained from the data product variance reported with the ungridded TROPOMI SIF. Our regression model explicitly models the measurement error, along with various sources of variation, which improves the estimation of the gridded SIF and allows for relatively smaller uncertainty estimates (standard errors) for the gridded SIF. However, since the TROPOMI uncertainty intervals are obtained from the TROPOMI data product variance directly, it includes the multiple sources of variation that we were able to separate from our standard error estimate. Additionally, our product uncertainties are based on OCO-2 observations which are higher in sample size than the available TROPOMI observations. Nevertheless, we note that the mean SIF derived from our product are consistent with those from TROPOMI, and are contained within the TROPOMI uncertainty bounds. 

\section{Conclusions} \label{sec:conclusions} 

This article presents a flexible Bayesian hierarchical regression framework for the estimation of gridded SIF and associated uncertainties from OCO-2 satellite observations with global coverage. The hierarchical structure of the regression model allows for convenient model specification due to the nested structure of the OCO-2 observations within gridded regions. Furthermore, the incorporation of a seasonal cycle term within the hierarchy allows us to leverage inherent seasonal patterns of vegetation for SIF estimation, thereby helping to inform both the gridded SIF estimates and their estimated uncertainties. This feature is particularly beneficial for regions and days with sparse data coverage, as the estimation is aided by information derived over time from the seasonal cycle. 

A significant contribution of this study lies in the comprehensive treatment of uncertainty estimation for the BHM gridded SIF product. The uncertainties are derived as standard errors of the estimated gridded SIF. Furthermore, we provide 95$\%$ credible intervals for the estimated gridded SIF as an alternative representation of the uncertainty of the SIF data product. These values are derived as the $2.5^{th}$ and $97.5^{th}$ quantiles of the posterior distribution of the $1 \degree$ SIF process, obtained using MCMC. Our regression framework provides a robust avenue to incorporate various sources of uncertainty and variation inherent to the estimation problem, including measurement errors of the OCO-2 SIF observations and variation of OCO-2 SIF within grid cells. By allowing these pieces to be modeled explicitly in the regression model, we aim to improve the estimation of the posterior distribution of the $1 \degree$ SIF process, and subsequently obtain reliable point estimates, standard errors and credible intervals of the daily, gridded SIF. This approach complements the estimation and UQ for the kriging-based gridded estimates of XCO$_2$ in the combined XCO$_2$ and SIF data product \citep{nguyen_meas_atbd_v4}. 

In this data product, we provide a combined SIF and XCO$_2$ product with uncertainties at the same $1 \degree \times 1 \degree$ target resolution and times. In doing so, our objective is to facilitate joint studies of XCO$_2$ and SIF at a global scale. Further, we use OCO-2 as the data source for the estimation of both the gridded SIF and XCO$_2$ to reduce bias in these joint studies. One major application which stands to benefit from the joint product is flux inversion studies. Our choice of $1 \degree \times 1 \degree$ target resolution accommodates the majority of global flux inversion models \citep{crowell_2019}. We note that the gridding resolution is relatively arbitrary, and the regression model can adapt to different target resolutions according to the application of interest. However, due to the seasonal cycle term in the hierarchical model, and 16 day temporal resolution of OCO-2 SIF, caution is warranted when there is insufficient temporal spread of sounding observation within a smaller resolution grid cell. Due to the revisit frequency of OCO-2, finer output grids may result in excessive temporal gaps in the observed data and lead to poor model performance. A data source such as TROPOMI, with approximately 1 day revisit frequencies, would yield improved model performance if finer outputs are desired.  

Looking forward, there are several additional avenues of interest, strengthened by the flexibility of the BHM framework. In the present model, we allow each grid cell to have its own seasonality term. However, a global model may benefit from allowing similar grid cells, such as those belonging to the same land cover or biome, to have a common, or at least correlated, seasonal cycle. This is strengthened by observing the distinct seasonal patterns in the distribution of SIF from the gridded product for each biome (Figures \ref{fig:boxplots_north}-\ref{fig:boxplots_south}). Our regression model allows for convenient estimation of the seasonal cycle terms through the posterior of the coefficients, which could be useful for tracking and quantifying changes in vegetation seasonality over the years. Additionally, while our focus has been on producing an OCO-2-based SIF product, future efforts could explore multiple data sources of SIF for a more comprehensive data fusion approach. The BHM approach can be adapted to utilize multiple data sources, which may improve estimation of the latent SIF process. Lastly, we utilize a single calendar year seasonality term in our modeling framework to allow for convenient estimation of gridded SIF by year. However, adjusting the start and end dates of the seasonality term could facilitate the development of a lower-latency product that can be updated more frequently. These advancements may lead to increased utility of SIF for global studies, and facilitate improved quantification and understanding of vegetation and carbon cycle dynamics.


\section*{Acknowledgements}
The research described in this paper
was performed at the Jet Propulsion Laboratory, California Institute of
Technology, under contract with NASA. Support was provided
by the Making Earth System Data Records for Use in Research Environments (MEaSUREs) program. The authors thank Russell Doughty, Thomas Kurosu, and Noel Cressie for helpful discussions. 

\bibliographystyle{ametsocV6}
\bibliography{sifrefs}

\begin{thebibliography}{35}
\providecommand{\natexlab}[1]{#1}
\providecommand{\url}[1]{\texttt{#1}}
\renewcommand{\UrlFont}{\rmfamily}
\providecommand{\urlprefix}{URL }
\expandafter\ifx\csname urlstyle\endcsname\relax
  \providecommand{\doi}[1]{https://doi.org/\discretionary{}{}{}#1}\else
  \providecommand{\doi}{https://doi.org/\discretionary{}{}{}\begingroup
  \urlstyle{rm}\Url}\fi
\providecommand{\eprint}[2][]{\url{#2}}

\bibitem[{Beer et~al.(2010)}]{beer_terrestrial_2010}
Beer, C., and Coauthors, 2010: Terrestrial gross carbon dioxide uptake:
  {Global} distribution and covariation with climate. \textit{Science},
  \textbf{329~(5993)}, 834--838, \doi{10.1126/science.1184984}.

\bibitem[{Berliner(1996)}]{berliner_hierarchical_1996}
Berliner, L.~M., 1996: Hierarchical {Bayesian} {Time} {Series} {Models}.
  \textit{Maximum {Entropy} and {Bayesian} {Methods}}, K.~M. Hanson, and R.~N.
  Silver, Eds., Springer Netherlands, Dordrecht, 15--22,
  \doi{10.1007/978-94-011-5430-7_3}.

\bibitem[{Bertolacci et~al.(2024)Bertolacci, Zammit-Mangion, Schuh, Bukosa,
  Fisher, Cao, Kaushik,, and Cressie}]{bertolacci_wombatv2_2024}
Bertolacci, M., A.~Zammit-Mangion, A.~Schuh, B.~Bukosa, J.~A. Fisher, Y.~Cao,
  A.~Kaushik, and N.~Cressie, 2024: Inferring changes to the global carbon
  cycle with {WOMBAT} v2.0, a hierarchical flux-inversion framework.
  \textit{The Annals of Applied Statistics}, \textbf{18~(1)},
  \doi{10.1214/23-AOAS1790}.

\bibitem[{Bloom et~al.(2020)}]{cardamom}
Bloom, A.~A., and Coauthors, 2020: Lagged effects regulate the inter-annual
  variability of the tropical carbon balance. \textit{Biogeosciences},
  \textbf{17~(24)}, 6393--6422, \doi{10.5194/bg-17-6393-2020}.

\bibitem[{Cressie and Wikle(2011)Cressie, and Wikle}]{cressie_wikle}
Cressie, N., and C.~K. Wikle, 2011: \textit{Statistics for Spatio-Temporal
  Data}. John Wiley \& Sons, Hoboken, NJ.

\bibitem[{Crowell et~al.(2019)}]{crowell_2019}
Crowell, S., and Coauthors, 2019: The 2015--2016 carbon cycle as seen from
  {OCO-2} and the global in situ network. \textit{Atmospheric Chemistry and
  Physics}, \textbf{19~(15)}, 9797--9831, \doi{10.5194/acp-19-9797-2019}.

\bibitem[{Dinerstein et~al.(2017)}]{biome}
Dinerstein, E., and Coauthors, 2017: An ecoregion-based approach to protecting
  half the terrestrial realm. \textit{BioScience}, \textbf{67~(6)}, 534--545,
  \doi{10.1093/biosci/bix014}.

\bibitem[{Doughty et~al.(2022)Doughty, Kurosu, Parazoo, Köhler, Wang, Sun,,
  and Frankenberg}]{doughty_ocosif_2022}
Doughty, R., T.~P. Kurosu, N.~Parazoo, P.~Köhler, Y.~Wang, Y.~Sun, and
  C.~Frankenberg, 2022: Global {GOSAT}, {OCO}-2, and {OCO}-3 solar-induced
  chlorophyll fluorescence datasets. \textit{Earth System Science Data},
  \textbf{14~(4)}, 1513--1529, \doi{10.5194/essd-14-1513-2022}.

\bibitem[{Frankenberg et~al.(2011)}]{frankenberg_grl_2011}
Frankenberg, C., and Coauthors, 2011: New global observations of the
  terrestrial carbon cycle from {GOSAT}: {Patterns} of plant fluorescence with
  gross primary productivity. \textit{Geophysical Research Letters},
  \textbf{38~(17)}, \doi{10.1029/2011GL048738}.

\bibitem[{Friedl and Sulla-Menashe(2015)Friedl, and Sulla-Menashe}]{MCD12C1}
Friedl, M., and D.~Sulla-Menashe, 2015: {MCD12C1 MODIS/Terra+Aqua} land cover
  type yearly {L3} global 0.05deg {CMG V006}. NASA EOSDIS Land Processes
  Distributed Active Archive Center, Dataset, \doi{10.5067/MODIS/MCD12C1.006}.

\bibitem[{Jacobson et~al.(2023)Jacobson, Cressie,, and
  Zammit-Mangion}]{jacobson_spatial_2023}
Jacobson, J., N.~Cressie, and A.~Zammit-Mangion, 2023: Spatial statistical
  prediction of solar-induced chlorophyll fluorescence ({SIF}) from
  multivariate {OCO}-2 data. \textit{Remote Sensing}, \textbf{15~(16)}, 4038,
  \doi{10.3390/rs15164038}.

\bibitem[{Joiner et~al.(2013)Joiner, Guanter, Lindstrot, Voigt, Vasilkov,
  Middleton, Huemmrich, Yoshida,, and Frankenberg}]{joiner_global_2013}
Joiner, J., L.~Guanter, R.~Lindstrot, M.~Voigt, A.~P. Vasilkov, E.~M.
  Middleton, K.~F. Huemmrich, Y.~Yoshida, and C.~Frankenberg, 2013: Global
  monitoring of terrestrial chlorophyll fluorescence from
  moderate-spectral-resolution near-infrared satellite measurements:
  methodology, simulations, and application to {GOME}-2. \textit{Atmospheric
  Measurement Techniques}, \textbf{6~(10)}, 2803--2823,
  \doi{10.5194/amt-6-2803-2013}.

\bibitem[{Joiner et~al.(2021)Joiner, Yoshida, Koehler, Frankenberg,, and
  Parazoo}]{joiner_sciamachy_2021}
Joiner, J., Y.~Yoshida, P.~Koehler, C.~Frankenberg, and N.~Parazoo, 2021: L2
  solar-induced fluorescence {(SIF)} from {SCIAMACHY}, 2003-2012. ORNL
  Distributed Active Archive Center, Dataset, \doi{10.3334/ORNLDAAC/1871}.

\bibitem[{Kleiber et~al.(2013)Kleiber, Katz,, and
  Rajagopalan}]{kleiber_daily_2013}
Kleiber, W., R.~W. Katz, and B.~Rajagopalan, 2013: Daily minimum and maximum
  temperature simulation over complex terrain. \textit{The Annals of Applied
  Statistics}, \textbf{7~(1)}, \doi{10.1214/12-AOAS602}.

\bibitem[{K\"{o}hler and Frankenberg(2020)K\"{o}hler, and
  Frankenberg}]{tropomi_dataset_caltech}
K\"{o}hler, P., and C.~Frankenberg, 2020: Ungridded {TROPOMI SIF} (at 740nm).
  CaltechDATA, Dataset, \doi{10.22002/D1.1347}.

\bibitem[{K\"{o}hler et~al.(2018)K\"{o}hler, Frankenberg, Magney, Guanter,
  Joiner,, and Landgraf}]{kohler_tropomi_2018}
K\"{o}hler, P., C.~Frankenberg, T.~S. Magney, L.~Guanter, J.~Joiner, and
  J.~Landgraf, 2018: Global retrievals of solar-induced chlorophyll
  fluorescence with {TROPOMI}: First results and intersensor comparison to
  {OCO-2}. \textit{Geophysical Research Letters}, \textbf{45}, 10,456--10,463,
  \doi{10.1029/2018GL079031}.

\bibitem[{Köhler et~al.(2020)Köhler, Behrenfeld, Landgraf, Joiner, Magney,,
  and Frankenberg}]{kohler_tropomi_2020}
Köhler, P., M.~J. Behrenfeld, J.~Landgraf, J.~Joiner, T.~S. Magney, and
  C.~Frankenberg, 2020: Global retrievals of solar-induced chlorophyll
  fluorescence at red wavelengths with tropomi. \textit{Geophysical Research
  Letters}, \textbf{47~(15)}, e2020GL087\,541, \doi{10.1029/2020GL087541}.

\bibitem[{Li and Xiao(2019)Li, and Xiao}]{li_gosif_2019}
Li, X., and J.~Xiao, 2019: A global, 0.05-degree product of solar-induced
  chlorophyll fluorescence derived from {OCO}-2, {MODIS}, and reanalysis data.
  \textit{Remote Sensing}, \textbf{11~(5)}, 517, \doi{10.3390/rs11050517}.

\bibitem[{Lindqvist et~al.(2015)}]{lindqvist_2015}
Lindqvist, H., and Coauthors, 2015: Does {GOSAT} capture the true seasonal
  cycle of carbon dioxide? \textit{Atmospheric Chemistry and Physics},
  \textbf{15~(22)}, 13\,023--13\,040, \doi{10.5194/acp-15-13023-2015}.

\bibitem[{MacBean et~al.(2022)}]{orchidas}
MacBean, N., and Coauthors, 2022: Quantifying and reducing uncertainty in
  global carbon cycle predictions: Lessons and perspectives from 15 years of
  data assimilation studies with the {ORCHIDEE} terrestrial biosphere model.
  \textit{Global Biogeochemical Cycles}, \textbf{36~(7)}, e2021GB007\,177,
  \doi{10.1029/2021GB007177}.

\bibitem[{Mohammed et~al.(2019)}]{mohammed_remote_2019}
Mohammed, G.~H., and Coauthors, 2019: Remote sensing of solar-induced
  chlorophyll fluorescence ({SIF}) in vegetation: 50 years of progress.
  \textit{Remote Sensing of Environment}, \textbf{231}, 111\,177,
  \doi{10.1016/j.rse.2019.04.030}.

\bibitem[{Nguyen et~al.(2024{\natexlab{a}})Nguyen, Johny, Liu, Kulawik, Baker,
  Hobbs, Braverman, Katzfuss,, and Yadav}]{measures_v4_sif}
Nguyen, H., M.~Johny, J.~Liu, S.~Kulawik, D.~Baker, J.~Hobbs, A.~Braverman,
  M.~Katzfuss, and V.~Yadav, 2024{\natexlab{a}}: {OCO-2} gridded bias-corrected
  {XCO2, SIF,} and other select fields aggregated as {Level 3} daily files.
  Goddard Earth Sciences Data and Information Services Center (GES DISC),
  Dataset, \doi{10.5067/0QR48EPN1BVR}.

\bibitem[{Nguyen et~al.(2024{\natexlab{b}})Nguyen, Liu, Kulawik, Baker, Hobbs,
  Johny, Braverman, Katzfuss,, and Yadav}]{nguyen_meas_atbd_v4}
Nguyen, H., J.~Liu, S.~Kulawik, D.~Baker, J.~Hobbs, M.~Johny, A.~Braverman,
  M.~Katzfuss, and V.~Yadav, 2024{\natexlab{b}}: {MEASURES} 2017 data fusion
  (v4) algorithm theoretical basis document. Pasadena, CA, URL:
  \url{https://docserver.gesdisc.eosdis.nasa.gov/public/project/MEaSUREs/XCO2_Data_Fusion/MEASURES_XCO2_ATBDdataVersion4.pdf}.

\bibitem[{Norton et~al.(2019)Norton, Rayner, Koffi, Scholze, Silver,, and
  Wang}]{bethy-scope}
Norton, A.~J., P.~J. Rayner, E.~N. Koffi, M.~Scholze, J.~D. Silver, and Y.-P.
  Wang, 2019: Estimating global gross primary productivity using chlorophyll
  fluorescence and a data assimilation system with the {BETHY-SCOPE} model.
  \textit{Biogeosciences}, \textbf{16~(15)}, 3069--3093,
  \doi{10.5194/bg-16-3069-2019}.

\bibitem[{{OCO-2/OCO-3 Science Team} et~al.(2020){OCO-2/OCO-3 Science Team},
  Payne,, and Chatterjee}]{ltsif_v11}
{OCO-2/OCO-3 Science Team}, V.~Payne, and A.~Chatterjee, 2020: {OCO-2} {Level}
  2 bias-corrected solar-induced fluorescence and other select fields from the
  {IMAP-DOAS} algorithm aggregated as daily files, retrospective processing
  {V11r}. Goddard Earth Sciences Data and Information Services Center {(GES
  DISC)}, Dataset, \doi{10.5067/OTRE7KQS8AU8}.

\bibitem[{Parazoo et~al.(2019)Parazoo, Frankenberg, Köhler, Joiner, Yoshida,
  Magney, Sun,, and Yadav}]{parazoo_jgr_2019}
Parazoo, N.~C., C.~Frankenberg, P.~Köhler, J.~Joiner, Y.~Yoshida, T.~Magney,
  Y.~Sun, and V.~Yadav, 2019: Towards a harmonized long-term spaceborne record
  of far-red solar-induced fluorescence. \textit{Journal of Geophysical
  Research: Biogeosciences}, \textbf{124~(8)}, 2518--2539,
  \doi{10.1029/2019JG005289}.

\bibitem[{Plummer(2017)}]{jags_manual}
Plummer, M., 2017: {JAGS} version 4.3.0 user manual. URL:
  \url{https://sourceforge.net/projects/mcmc-jags/files/Manuals/4.x/jags_user_manual.pdf}.

\bibitem[{Poppick and McKinnon(2020)Poppick, and
  McKinnon}]{poppick_quantreg_2020}
Poppick, A., and K.~A. McKinnon, 2020: Observation-based simulations of
  humidity and temperature using quantile regression. \textit{Journal of
  Climate}, \textbf{33~(24)}, 10\,691--10\,706, \doi{10.1175/JCLI-D-20-0403.1}.

\bibitem[{Shiga et~al.(2018)Shiga, Tadić, Qiu, Yadav, Andrews, Berry,, and
  Michalak}]{shiga_2017}
Shiga, Y.~P., J.~M. Tadić, X.~Qiu, V.~Yadav, A.~E. Andrews, J.~A. Berry, and
  A.~M. Michalak, 2018: Atmospheric {CO$_2$} observations reveal strong
  correlation between regional net biospheric carbon uptake and solar-induced
  chlorophyll fluorescence. \textit{Geophysical Research Letters},
  \textbf{45~(2)}, 1122--1132, \doi{10.1002/2017GL076630}.

\bibitem[{Su and Yajima(2015)Su, and Yajima}]{r2jags}
Su, Y.-S., and M.~Yajima, 2015: R2jags: Using {R} to run {‘JAGS’}.
  \textit{R package version 0.5-7}, \textbf{34}.

\bibitem[{Sun et~al.(2018)Sun, Frankenberg, Jung, Joiner, Guanter, Köhler,,
  and Magney}]{sun_oco2_2018}
Sun, Y., C.~Frankenberg, M.~Jung, J.~Joiner, L.~Guanter, P.~Köhler, and
  T.~Magney, 2018: Overview of solar-induced chlorophyll fluorescence {(SIF)}
  from the {Orbiting Carbon Observatory-2}: Retrieval, cross-mission
  comparison, and global monitoring for {GPP}. \textit{Remote Sensing of
  Environment}, \textbf{209}, 808--823, \doi{10.1016/j.rse.2018.02.016}.

\bibitem[{Xu et~al.(2021)}]{xu_2021}
Xu, S., and Coauthors, 2021: Structural and photosynthetic dynamics mediate the
  response of sif to water stress in a potato crop. \textit{Remote Sensing of
  Environment}, \textbf{263}, 112\,555, \doi{10.1016/j.rse.2021.112555}.

\bibitem[{Yao et~al.(2022)}]{sif_retrieval_yao}
Yao, L., and Coauthors, 2022: Retrieval of solar-induced chlorophyll
  fluorescence {(SIF)} from satellite measurements: comparison of {SIF between
  TanSat and OCO-2}. \textit{Atmospheric Measurement Techniques},
  \textbf{15~(7)}, 2125--2137, \doi{10.5194/amt-15-2125-2022}.

\bibitem[{Zhang et~al.(2023)}]{zhang_sifinv_2023}
Zhang, M., and Coauthors, 2023: Solar-induced fluorescence helps constrain
  global patterns in net biosphere exchange, as estimated using atmospheric
  {CO$_2$} observations. \textit{Journal of Geophysical Research:
  Biogeosciences}, \textbf{128~(12)}, e2023JG007\,703,
  \doi{10.1029/2023JG007703}.

\bibitem[{Zhang et~al.(2018)Zhang, Joiner, Alemohammad, Zhou,, and
  Gentine}]{zhang_csif_2018}
Zhang, Y., J.~Joiner, S.~H. Alemohammad, S.~Zhou, and P.~Gentine, 2018: A
  global spatially contiguous solar-induced fluorescence ({CSIF}) dataset using
  neural networks. \textit{Biogeosciences}, \textbf{15~(19)}, 5779--5800,
  \doi{10.5194/bg-15-5779-2018}.

\end{thebibliography}

\end{document}